\documentclass[preprint]{aastex}
\usepackage{gensymb}

\shorttitle{Chemical Imaging of CO snow line}
\shortauthors{}

\begin{document}

\title{Chemical Imaging of the CO Snow Line in the HD 163296 Disk}

\author{Chunhua~Qi\altaffilmark{1}, 
Karin~I.~\"Oberg\altaffilmark{1}, 
Sean~M.~Andrews\altaffilmark{1},
David~J.~Wilner\altaffilmark{1},
Edwin~A.~Bergin\altaffilmark{2},
A.~Meredith~Hughes\altaffilmark{3},
Michiel~Hogherheijde\altaffilmark{4},
Paola~D'Alessio\altaffilmark{5,6}}

\altaffiltext{1}{Harvard-Smithsonian Center for Astrophysics, 60 Garden Street, 
Cambridge, MA 02138, USA} 

\altaffiltext{2}{Department of Astronomy, University of Michigan, 500 Church
  Street, Ann Arbor, MI 48109, USA}

\altaffiltext{3}{Van Vleck Observatory, Astronomy Department, Wesleyan
  University, 96 Foss Hill Drive, Middletown, CT 06459, USA}

\altaffiltext{4}{Leiden Observatory, Leiden University, PO Box 9513, 2300 RA
  Leiden, The Netherlands}

\altaffiltext{5}{Centro de Radioastronom\'{i}a y Astrof\'{i}sica, 
Universidad Nacional Aut\'{o}noma de M\'{e}xico, 58089 Morelia, 
Michoac\'{a}n, M\'{e}xico}

\altaffiltext{6}{Deceased Nov 14, 2013}

\begin{abstract}
The condensation fronts (snow lines) of H$_2$O, CO and other abundant
volatiles in the midplane of a protoplanetary disk affect several
aspects of planet formation. Locating the CO snow line, where the CO
gas column density is expected to drop substantially, based solely on
CO emission profiles is challenging. This has prompted an exploration of 
chemical signatures of CO freeze-out. 
We present ALMA Cycle 1 observations of the N$_2$H$^+$~$J=3-2$ and 
DCO$^+$~$J=4-3$ emission lines toward the disk around the Herbig Ae star 
HD~163296 at $\sim0.5''$ (60 AU) resolution, and evaluate their
utility as tracers of the CO snow line location.  The N$_2$H$^+$
emission is distributed in a ring with an inner radius at 90~AU,
corresponding to a midplane temperature of 25~K. 
This result is consistent with a new analysis of optically thin
C$^{18}$O data, which implies a sharp drop in CO abundance at 90 AU.
Thus N$_2$H$^+$ appears to be a robust tracer of the
midplane CO snow line. The DCO$^+$ emission also has a ring
morphology, but neither the inner nor the outer radius coincides with
the CO snow line location of 90~AU, indicative of a complex
relationship between DCO$^+$ emission and CO freeze-out in the disk
midplane. Compared to TW~Hya, CO freezes  out at a higher temperature
in the disk around HD~163296 (25 vs. 17~K in the TW Hya disk), perhaps
due to different ice compositions. This highlights the importance of
actually measuring the CO snow line location, rather than assuming a
constant CO freeze-out temperature for all disks.    
\end{abstract}


\section{Introduction}

A condensation front is the two-dimensional surface in a protoplanetary 
disk where abundant volatiles freeze out of the gas phase onto solid 
particles: the ``snow line" marks this condensation front at the disk 
midplane.  Based on observations of protostars and comets, as well as 
theoretical models of disk chemistry, the most important snow lines are 
due to H$_2$O, CO$_2$, CO and N$_2$ freeze-out
\citep{Oberg11c,Mumma11}. At each of these snow lines there will be a
substantial increase in 
particle size and solid surface density, augmented by cold finger and
pressure trap effects, which may speed up planetesimal formation  
  \citep{Ciesla06,Johansen07,Chiang10,Gundlach11,Ros13}. Freeze-out of
  different volatiles will also affect grain stickiness, another key
  factor for particle growth in disks. The importance of an individual
  snow line on the planet formation process in a disk will depend on
  its radius with respect to the distribution of mass in the disk,
  i.e. the density profile. To evaluate the importance of different
  snow lines to planet formation therefore requires precise
  constraints on snow line locations. Snow line locations are also
  important for setting the bulk composition of planets. That is,
  where a planet accretes its core and envelope with respect to
  different snow lines will affect its solid and atmospheric C/O ratio,
  and therefore the planetary chemistry \citep{Oberg11d}.

The locations of snow lines depend on the volatile composition
(e.g. whether most nitrogen is in N$_2$ or NH$_3$), the balance between
freeze-out and thermal and non-thermal desorption rates (which are themselves 
set by a combination of density, temperature, and radiation fields),
and dynamical processes in the disk, including drift of grains and gas
diffusion  \citep{Oberg11b,Oka12,Ali-Dib14,Baillie15}.  
Exact snow line locations are therefore difficult to predict theoretically;  
observational constraints are key. 
Of the major snow lines, the CO one should be the most straightforward
to constrain observationally using millimeter interferometry.  
CO is the  most abundant molecule after H$_2$ and its snow line is
theoretically expected in the outer disk, where the  
temperature reaches $\sim$20~K. In the disks around T Tauri 
and Herbig Ae stars, this threshold is at a radius of $\sim$20--150~AU, and can 
thus in principle be resolved with modern millimeter interferometers. In
practice, the presence of warmer CO in the gas phase at larger heights above 
the midplane (in the disk atmosphere) at all radii makes it difficult to 
directly locate the CO snow line.  

The most detailed attempt to date to use observations of CO emission lines to 
infer the CO snow line location was carried out by \citet{Qi11} for the disk 
around HD~163296 at a distance of $\sim$122 pc, 
the target of this study. \citet{Qi11} used the spectral 
energy distribution (SED) along with multiple optically thick $^{12}$CO 
transitions to constrain the two-dimensional (radial and vertical) density and 
temperature structure of the disk. The CO snow line location was then 
constrained using the less optically thick $^{13}$CO $J$=$2-1$ line emission, 
and found to be at a radius of 155~AU. 

Another approach to constrain the CO snow line location is to
exploit chemical effects that are regulated by CO freeze-out \citep{Qi13a}. 
One such method of tracing the CO snow line uses the ion N$_2$H$^+$
\citep{Qi13c}.  CO in the gas phase will significantly slow down N$_2$H$^+$ 
formation {\it and} speed up N$_2$H$^+$ destruction \citep{Bergin01}.  
The predicted anti-correlation of gas-phase CO and N$_2$H$^+$ has been
confirmed by observations in pre-stellar and protostellar environments
\citep{Caselli99,Bergin02,Jorgensen04}. 
 In disks, such anti-correlation should result in a N$_2$H$^+$ ring,
 where the inner ring 
 radius traces the onset of CO freeze-out  (i.e. CO depletion from the
 gas-phase) at the CO snow line. 
 \citet{Qi13c} imaged
N$_2$H$^+$ emission from the disk around the T Tauri star TW~Hya and found a 
large ring with an inner edge at 30~AU.  Associating that inner ring location 
with the CO snow line, a CO freeze-out temperature of 17~K was extracted based 
on a detailed model for the disk density and temperature structure. 

\citet{Mathews13} advocate another tracer based on the distribution 
of the ion DCO$^+$.  In most chemical models, the dominant DCO$^+$ formation 
pathway involves H$_2$D$^+$, which promotes an increasing DCO$^+$ abundance 
at low ($<$30~K) temperatures as long as CO is available in the gas phase.  
Therefore, the DCO$^+$ emission in a disk is expected to peak just {\it 
interior} to the CO snow line.  \citet{Mathews13} imaged DCO$^+$ emission 
toward the HD~163296 disk and found a ring morphology with an outer edge that
coincides with the \citet{Qi11} CO snow line of 155~AU, supporting this
theory.  However, more recent models suggest that other, warmer formation
pathways of DCO$^+$ may be also important in disks, increasing the
DCO$^+$ abundance in higher (warmer) disk layers and thus weakening
the relationship between the DCO$^+$ abundance distribution and the
midplane CO snow line \citep{Favre15}.  

In this article, we revisit the CO snow line in the disk around HD 163296.
The aim is to determine the CO snow line location 
using observations of N$_2$H$^+$ emission. We also present 
archival C$^{18}$O and new DCO$^+$ observations and compare the constraints 
provided by these different chemical tracers of the gas-phase distribution of 
CO.  These data and their calibration are described in \S\ref{sec:obs}.  We 
present the N$_2$H$^+$, DCO$^+$, and C$^{18}$O emissions maps from the 
HD~163296 disk, along with their associated abundance profiles derived from 
models, in \S\ref{sec:res}.  The results are used to determine the CO snow line 
locations in this disk, and to evaluate the utility and robustness of these 
different probes, in \S\ref{sec:disc}.  

\begin{table*}[ht]
\caption{Observational Parameters$^a$}
\begin{tabular}{lcc}
\hline\hline
\multicolumn{3}{c}{Project 2012.1.00681.S}\\
\hline
Continuum (GHz) & \multicolumn{2}{c}{279.5} \\
Beam Size (FWHM) & \multicolumn{2}{c}{$0.''46 \times 0.''35$} \\
P.A. (\degree) & \multicolumn{2}{c}{$-$83.3} \\
RMS Noise (mJy\,beam$^{-1}$) & \multicolumn{2}{c}{0.19} \\
Flux (mJy) & \multicolumn{2}{c}{719$\pm$72} \\
\hline
Lines & N$_2$H$^+$ $J=3-2$ & DCO$^+$ $J=4-3$ \\
Rest frequency (GHz) & 279.512 & 288.144\\
Beam Size (FWHM) & $0.''75 \times 0.''60$ & $0.''58 \times 0.''37$ \\
P.A. (\degree) & $-$75.4 & $-$73.2 \\
Channel spacing (km\,s$^{-1}$) & 0.13 & 0.063 \\
Integrated Flux (Jy\,km\,s$^{-1}$) & 0.52$\pm$0.05 & 1.34$\pm$0.13 \\
\hline\hline
\multicolumn{3}{c}{Project 2011.1.00010.SV}\\
\hline
Continuum (GHz) & \multicolumn{2}{c}{218.3} \\
Beam Size (FWHM) & \multicolumn{2}{c}{$0.''73 \times 0.''57$} \\
P.A. (\degree) & \multicolumn{2}{c}{73.2} \\
RMS Noise (mJy\,beam$^{-1}$) & \multicolumn{2}{c}{0.19} \\
Flux (mJy) & \multicolumn{2}{c}{609$\pm$61} \\
\hline
Lines & C$^{18}$O $J=2-1$  & $^{13}$CO $J=2-1$ \\
Rest frequency (GHz) & 219.560 & 220.399\\
Beam Size (FWHM) & $0.''89 \times 0.''72$ & $0.''87 \times 0.''70$ \\
P.A. (\degree) & 82.8 & 84.3 \\
Channel spacing (km\,s$^{-1}$) & 0.33 & 0.33 \\
Integrated Flux (Jy\,km\,s$^{-1}$) & 6.8$\pm$0.7 & 19$\pm$2 \\
\hline
\end{tabular}
\tablenotetext{a}{All quoted values assume natural weighting.}
\label{tab:obs}
\end{table*}

\section{Observations\label{sec:obs}}

The observations of N$_2$H$^+$ and DCO$^+$ from the HD~163296 disk were
carried out with the Atacama Large Millimeter/submillimeter Array (ALMA) on 
21 May 2014, with 32 12~m antennas (ALMA Cycle 1 project 2012.1.00681.S). The 
total integration time on HD~163296 was 45 minutes. Baselines ranged from 
18--650 m (17--605 k$\lambda$).   

The correlator was configured to observe four spectral windows (SPWs): two 
were centered at the rest frequencies of the N$_2$H$^+$ $J=3-2$ and 
DCO$^+$ $J=4-3$ transitions (see Table 1), one contained a H$_2$CO
line\footnote{We defer the analysis of H$_2$CO to future work because 
the emission distribution can be complicated with respect
to the location of the CO snow line \citep[e.g.][]{Loomis15}.}, and  
the last SPW sampled only the continuum.  The nearby quasar J1733-1304 was
used for phase and gain 
calibration with a mean flux density of 1.048 Jy. The visibility data
were reduced and calibrated in CASA 4.2.

The continuum visibilities were extracted by averaging the line-free
channels. We carried out self-calibration procedures on the 
continuum as demonstrated in the HD~163296 Science
Verification Band 7 CASA Guides, 
which are available online\footnote{
https://almascience.nrao.edu/almadata/sciver/HD163296Band7}.  
We applied the continuum self-calibration 
correction to
the N$_2$H$^+$~$J=3-2$ and DCO$^+$~$J=4-3$ line data and subtracted
the continuum emission in the visibility domain. 

To optimize the signal-to-noise ratio of 
the N$_2$H$^+$ channel maps, we applied a Gaussian taper to the visibilities 
that corresponds to 0.5\arcsec ~FWHM in the image domain. The
resulting synthesized beam for the N$_2$H$^+$ $J=3-2$ data cube is 
$0.''75 \times 0.''60$, slightly larger than the beam used to observe
DCO$^+$~$J=4-3$, ($0.''58 \times 0.''37$). The integrated fluxes and the beam
information are reported in Table~\ref{tab:obs}. We note that this
tapering does not affect the model fit procedure that is used to
extract the radial abundance profiles of the two species, since that fitting
is done in the visibility domain. 

We also make use of the $^{13}$CO and C$^{18}$O~$J=2-1$ emission
toward HD~163296 in the ALMA science verification (SV) program
2011.0.00010.SV. See the observations and data reduction details in   
\cite{Rosenfeld13}. The continuum and line intensities are also
reported in Table~\ref{tab:obs} and agree with measurements in
\cite{Rosenfeld13}. 

\begin{figure*}[ht]
\epsscale{1}
\plotone{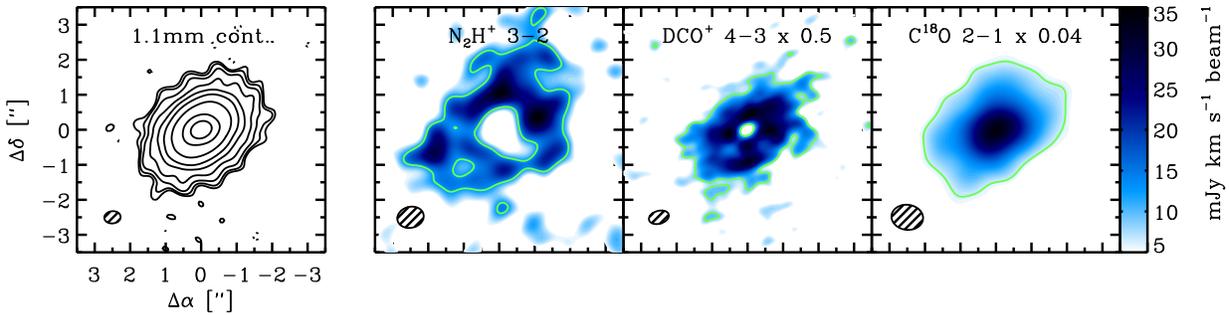}
\caption{ Continuum and integrated intensity images of
  N$_2$H$^+$~$J=3-2$, DCO$^+$~$J=4-3$ and C$^{18}$O~$J=2-1$ toward the
  disk around 
  HD~163296. The 
  contours in the continuum image are 2$\sigma$+[2,4,8,16,...]$\sigma$
  with the continuum $\sigma=0.19$
  mJy~beam$^{-1}$. The green contours in the line images are 3$\sigma$
  except 15$\sigma$ for C$^{18}$O~$J=2-1$.  
The integrated line emission scale is shown to the
  right of the C$^{18}$O panel. Synthesized beams are drawn in the
  bottom left corner of each panel. \label{fig:hdmaps}} 
\end{figure*} 

\section{Results}\label{sec:res}

Figure~\ref{fig:hdmaps} shows the continuum and integrated intensity
maps of N$_2$H$^+$~$J=3-2$, DCO$^+$~$J=4-3$ and C$^{18}$O from the HD~163296
disk. The channel maps used to construct those maps are also shown in the
Appendix. The 1.1~mm continuum image in Figure~\ref{fig:hdmaps} shows
centrally peaked emission with an integrated flux density of
719$\pm$72 mJy (RMS noise level of  0.19 mJy~beam$^{-1}$), consistent
with previous SMA measurements \citep{Qi13a}. 
The integrated intensity of N$_2$H$^+$~$J=3-2$, 0.52$\pm$0.05 Jy~km~s$^{-1}$, 
also agrees well with previous SMA observations \citep{Qi13a}.  The N$_2$H$^+$ 
emission is clearly distributed in a ring (disk inclination at around
44$\degree$, see Qi et al. 2011) with an approximate inner
edge radius of 0.6 to 0.8$''$ (based on visual inspection). The
DCO$^+$ emission is also distributed in a ring, albeit  
with a substantially smaller inner edge of 0.2 to 0.4$''$. The DCO$^+$ 
emission is also considerably more compact than the N$_2$H$^+$
emission. The C$^{18}$O emission shows centrally peaked emission
  around the 1.1~mm continuum peak.

In the following sections, we use the spatially and spectrally resolved 
N$_2$H$^+$, DCO$^+$, and C$^{18}$O line emission to constrain the abundance 
profiles of the three molecules, and their relation to one another and the CO 
snow line.

\subsection{N$_2$H$^+$ and the CO snow line location}

\begin{table*}
\caption{Fitting results$^a$}
\begin{center}
\begin{tabular}{l c c c c c}
\hline
\hline
Transitions & R$_{in}$ (AU) & $p$ &
N$_{100}$ (cm$^{-2}$) & R$_{out}$ (AU) & $\sigma_s$,$\sigma_m$$^b$ \\
\hline
N$_2$H$^+$ $3-2$ & 90$^{+8}_{-6}$ &  $-$0.7$^{+0.1}_{-0.2}$   &
(2.6$\pm$0.1) $\times$10$^{11}$ & 500$^b$ & 1,32  \\
DCO$^+$ $4-3$ & 40$^{+6}_{-3}$ &  $-$1.0$^{+0.1}_{-0.1}$  & (8.5$\pm$0.5) $\times$10$^{11}$ & 290$^{+6}_{-8}$ & 1,32 \\
\hline
$^{a}$Errors within 3$\sigma$.\\
$^{b}$Fixed parameters.\\
\end{tabular}
\end{center}
\label{tab:models}
\end{table*}   

\begin{figure*}[htb]
\epsscale{1}
\plotone{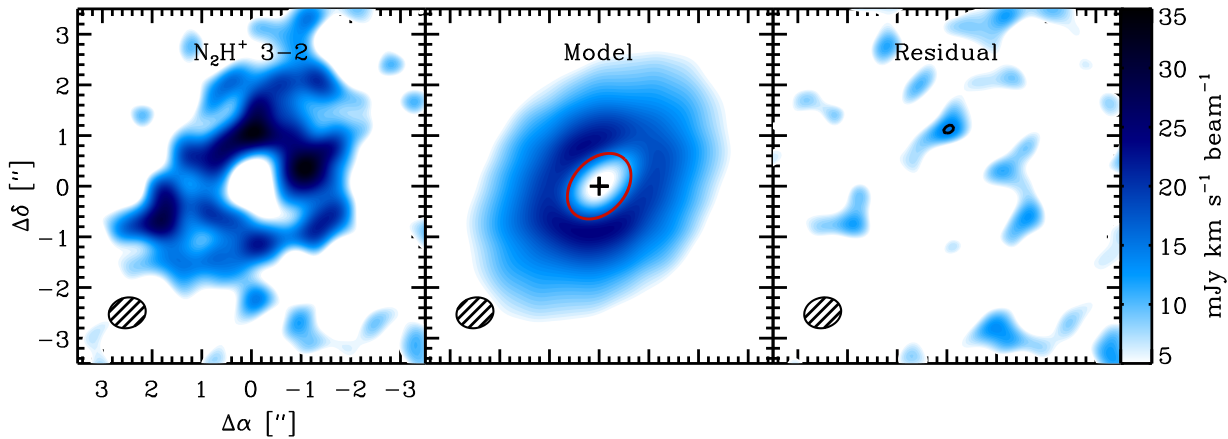}
\caption{ 
N$_2$H$^+$ $3-2$ observations, simulated
  observations of the best-fit N$_2$H$^+$ model, and the imaged
  residuals, calculated from the visibilities. The red ellipse marks
  the best-fit inner radius of the N$_2$H$^+$ ring at 90 AU, the CO
  snow line in the disk midplane. The residual shows also the contours
  in steps of 3$\sigma$. 
The integrated line emission scale is shown to the
  right of the upper right panel. Synthesized beams are drawn in the
  bottom left corner of each panel.  \label{fig:n2hp32mom}} 
\end{figure*}

Based on the simple chemical connection between N$_2$H$^+$ abundance and CO 
depletion, we expect that N$_2$H$^+$ emission provides the strongest available
constraints on the CO snow line location. We extract the N$_2$H$^+$
abundance profile in the HD~163296 disk following the methodology of 
\citet{Qi11,Qi13c}. We adopt the density and temperature disk
structure from \citet{Qi11}, which constitutes a physically
self-consistent accretion disk model with an exponentially tapered 
edge that matches the HD~163296 spectral energy distribution and
optically thick multi-transition $^{12}$CO observations.
The N$_2$H$^+$ column density profile is parameterized as
$N_{100}\times(r/100)^p$ with 
an inner and outer boundary, where 
$N_{100}$ is the column density at 
100 AU in cm$^{-2}$, $r$ is the distance from the star in AU, and $p$
is a power-law index. We fit for
an inner radius $R_{in}$ and outer radius $R_{out}$ together with
the power-law parameters, $N_{100}$ and $p$.  

In the vertical dimension, the N$_2$H$^+$ abundance is assumed to be constant 
between the disk surface ($\sigma_s$) and midplane ($\sigma_m$) boundaries at 
each $r$. These boundaries are described in terms of $\Sigma_{21}=\Sigma_H / 
(1.59\times10^{21} {\rm cm}^{-2})$, where $\Sigma_H$ is the hydrogen column 
density (measured downward from the disk surface) in the adopted physical 
model. This simple model approach approximates the outcome of disk chemistry 
models, which find that most molecules are present in such a layered 
structure. {We fix the vertical boundary values $\sigma_s$,
  $\sigma_m$ to 1, 32, respectively, as listed in
  Table~\ref{tab:models}, based on the chemical models of
  \citet[Fig. 8]{Aikawa06}, which shows that the abundances of the
common disk molecules HCN and HCO+ are sharply reduced  at
$\Sigma_{21}<1$ (the disk surface) and $\Sigma_{21}>10$ (due to
freeze-out in the midplane). Because DCO$^+$ and N$_2$H$^+$ production is
enhanced at low temperatures, we expect both molecules to be present
deeper toward the midplane than HCN and
HCO$^+$~\citep[e.g.][]{Willacy07, Walsh10}. So we choose a midplane
boundary $\sigma_m$ of 32, the logarithmic mean of 10 and 100.
We find that, given the 
current sensitivity of the data and moderate
inclination (44$\degree$) of the disk, the fitting results on the
radial boundary  
parameters $R_{in}$ and $R_{out}$ are not sensitive to the choice of these 
vertical boundaries.  
Following an initial model optimization, we then
also fixed R$_{out}$ to 500~AU~\citep[the extent of the CO
  disk,][]{Qi11}, since the data provides no constraints on the
N$_2$H$^+$ beyond a few hundred AU due to a lack of sensitivity. 

The best-fit parameter estimates are obtained by minimizing
$\chi^2$, the weighted residual of the complex visibilities 
measured at the ($u,v$)-spacings sampled by ALMA.
We use the two-dimensional Monte Carlo software RATRAN \citep{Hogerheijde00} 
to calculate the radiative transfer and molecular excitation. 
The collisional rates are adopted from \citet{Flower99} based on
HCO$^+$ collisional rates with H$_2$ and the
molecular data files are retrieved from the Leiden Atomic and Molecular 
Database \citep{Schoier05}. 

Figure~\ref{fig:n2hp32mom} compares the observed
N$_2$H$^+$ emission map with the best-fit model
(Table~\ref{tab:models}). See Table 3 of \citet{Qi11} for other
related disk model parameters including disk inclination and
orientation. The best-fit model matches the observations very well: the
residual image shows no significant emission above the 3$\sigma$
level. Based on this model, the N$_2$H$^+$ column density is
2.6$\pm$0.1 $\times$10$^{11}$ cm$^{-2}$ at 100 AU and the power-law
index is $-0.7$.  The most important parameter in the model fit is
$R_{in}$, since it is associated with the CO snow line location. We
find $R_{in}$=90$^{+8}_{-6}$ AU.  Note that this radius is 65~AU interior to 
the CO snow line location estimated by \citet{Qi11}. The N$_2$H$^+$ inner edge 
corresponds to a midplane temperature of 25~K in the underlying model of the 
disk structure.  

\subsection{C$^{18}$O constraints on the CO snow line}

\begin{figure*}[htb]
\includegraphics[width=5.3in]{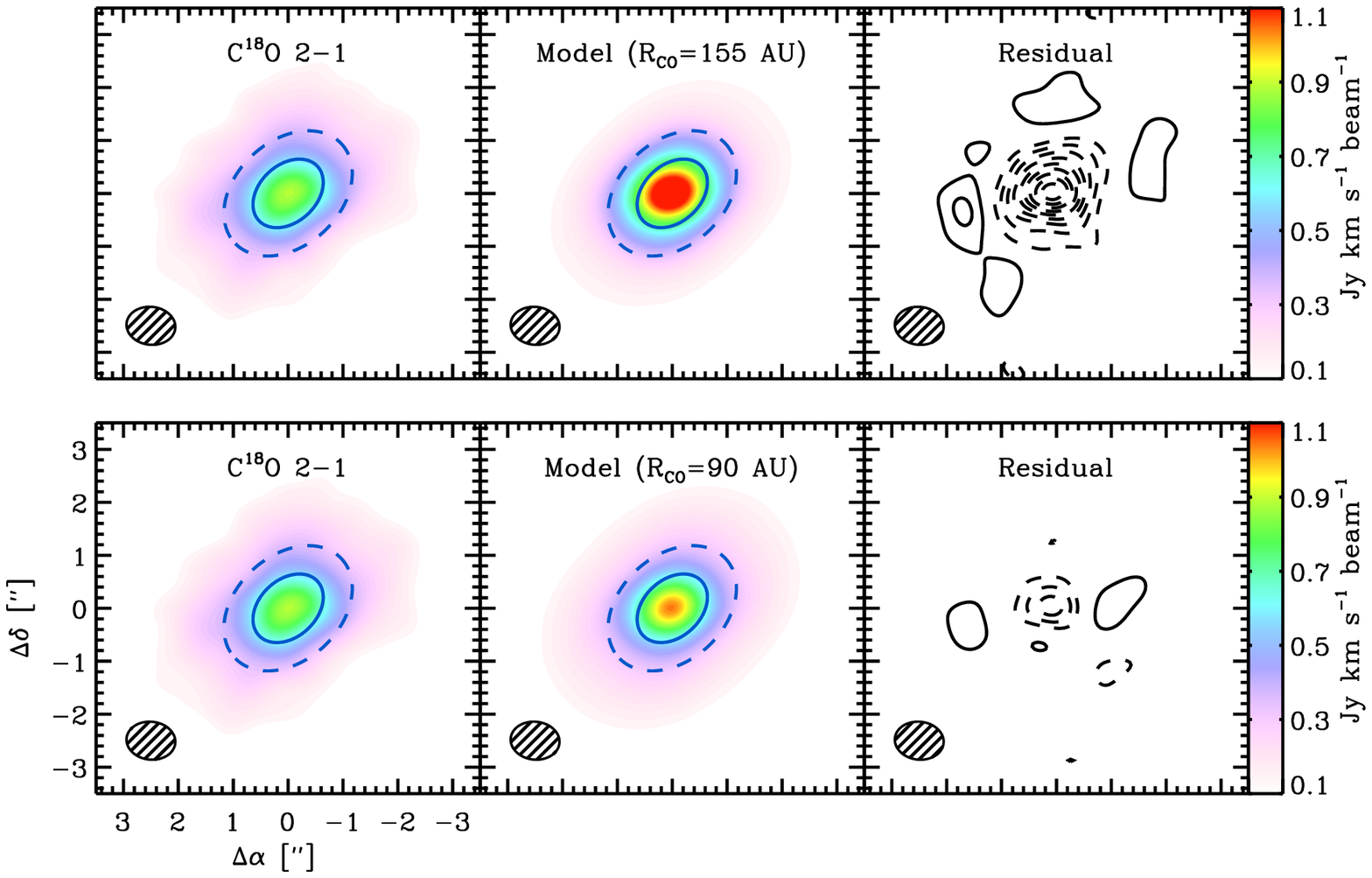}
\vskip 5mm
\includegraphics[width=5.3in]{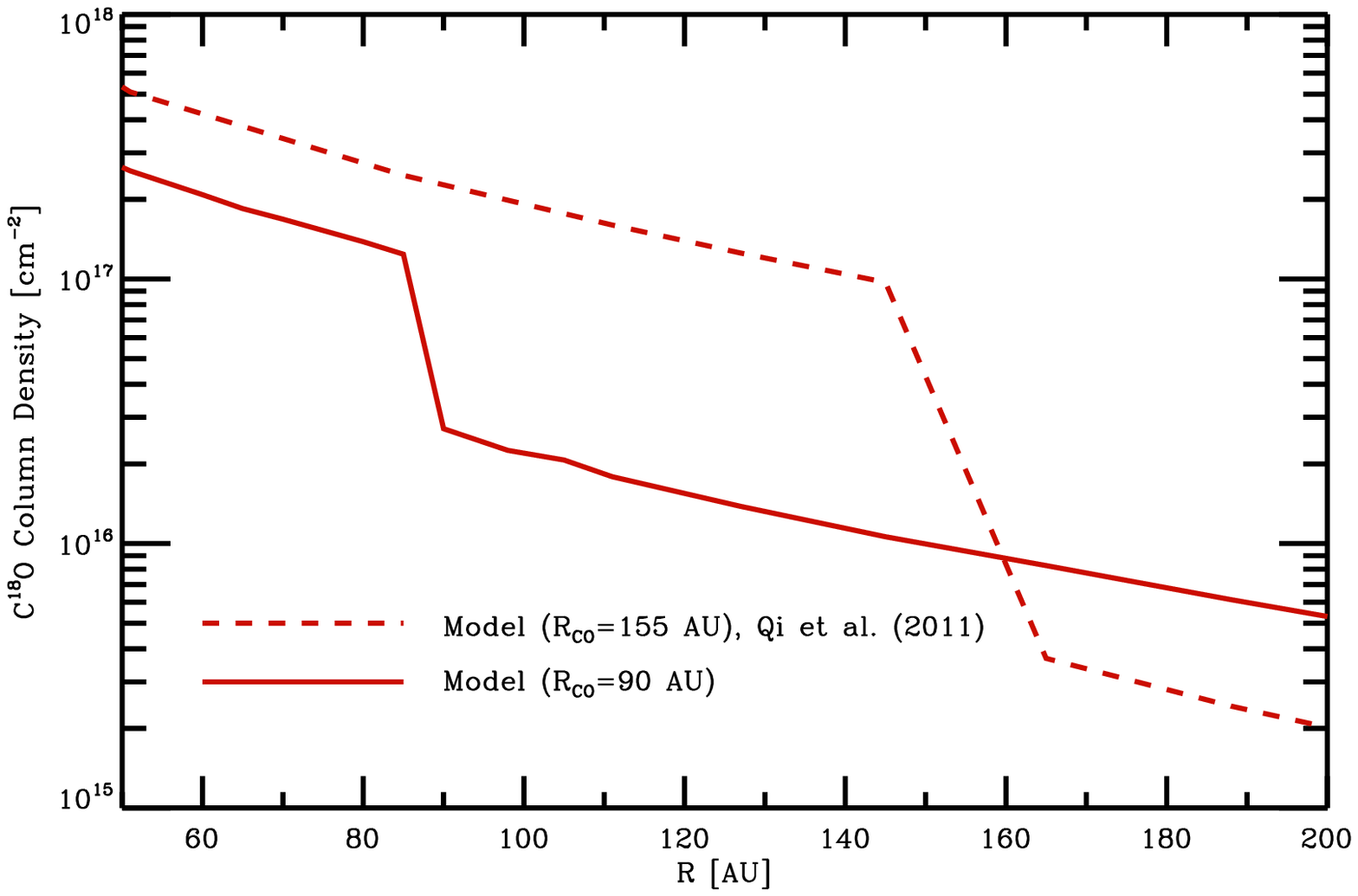}
\caption{Model results for the C$^{18}$O abundance structure toward
  HD 163296. {\it Top panels:} Observations of C$^{18}$O 2--1 toward HD~162396 (left)
  together with the modified model based on Qi et al. (2011) with
  T$_{\rm CO}$=19 K and R$_{\rm CO}$=155 AU (upper middle panel) and the
  best-fit model with T$_{\rm CO}$=25 K, R$_{\rm CO}$=90 AU (lower
  middle panel). The residuals show the contours
  in steps of 3$\sigma$. Solid and dashed blue ellipses mark the radii of 90 AU
  and 155 AU, respectively. The integrated line
  emission scale is shown to the right of the panels
  Synthesized beams are drawn in the bottom left corner of each
  panel. {\it Bottom panel:} C$^{18}$O column density
  profiles for the two models.  
\label{fig:c18omodels}}
\end{figure*}

The CO snow line location that would be implied by the N$_2$H$^+$ inner ring 
edge is in clear conflict with the value derived by \citet{Qi11}, which was 
based on modeling the $^{13}$CO line emission.  Here we (re-)analyze the 
spatially resolved CO isotopologue emission from the HD~163296 disk using the
same model framework as outlined above and ALMA archival $^{13}$CO and
C$^{18}$O data. 

As a first step we repeated the analysis presented in \citet{Qi11}
using the new $^{13}$CO $J$=2$-$1 data from ALMA, with substantially improved 
sensitivity (\citet{Qi11} used Submillimeter Array observations). Doing so, we 
recover the best-fit model presented by \citet{Qi11}, with a CO snow line at
155~AU. Next, we tried to fit the new ALMA C$^{18}$O $J$=2$-$1 data using the 
same model structure and CO snow line location, but with the fractional 
abundance of C$^{18}$O as a free parameter.  Figure~\ref{fig:c18omodels} (top panels) demonstrates that such a model cannot fit the C$^{18}$O 
data very well. The model with a 155~AU snow line systematically over-predicts 
the C$^{18}$O emission interior to 155~AU and under-predicts it at larger 
radii, regardless of the level of CO depletion, indicating that there is no 
sharp drop in the CO abundance at 155~AU.  Compared to $^{13}$CO, C$^{18}$O 
should be less affected by opacity effects and therefore provide a more direct
constraint on the CO column density profile.  

The failure of the model to reproduce the C$^{18}$O emission morphology with a 
fixed 155~AU snow line suggests that $^{13}$CO emission is not a robust tracer 
of CO depletion by freeze-out. One possible explanation of this
discrepancy is 
that the $^{13}$CO line is optically thick out to 
$\sim$155\,AU; the apparent 155~AU snow line inferred by \citet{Qi11} actually
reflects the transition to optically thin emission, with a pronounced intensity 
drop without a corresponding column density decrease.  In order to maintain 
high $^{13}$CO optical depths at such large radii, there must be some
midplane gas-phase CO abundance exterior to the CO snow line. \citet{Qi11} 
assumed {\it complete} freeze-out of CO whenever the temperature is below a 
critical value.  However, non-thermal desorption mechanisms (e.g., UV 
photodesorption) can maintain a relatively large CO fraction in the gas phase 
\citep{Oberg08b} at the low densities present in the outer disk. 

The C$^{18}$O isotopologue is expected to be $\sim$8$\times$ less 
abundant than $^{13}$CO, and therefore its emission morphology is much less 
likely to be affected by these opacity effects.  To explore its ability to 
constrain the CO snow line, we repeated the analysis methodology described by 
\citet{Qi11}, but now with both the CO freeze-out temperature (T$_{\rm CO}$) 
and the CO freeze-out fraction (depletion factor; $F_{\rm CO}$) as free 
parameters (while fixing the surface boundary $\sigma_s$ as 0.79 to
keep the problem  
computationally tractable). The lower boundary (toward midplane) is
still governed by T$_{\rm CO}$, and the abundance of CO drops
substantially when T$<$T$_{\rm CO}$ but not to 0, i.e. no complete
freeze-out of CO. 
We find a best-fit T$_{\rm CO} = 25-26$ K, which 
occurs at a radius of 85--90 AU for the adopted disk structure. The
best-fit C$^{18}$O abundance is 9$\times$10$^{-8}$, corresponding to
CO abundance of 5$\times$10$^{-5}$ assuming an abundance ratio of 
CO/C$^{18}$O=550,  and the depletion factor is about 5. The depletion
factor depends on the detailed thermal and non-thermal desorption processes
of CO ice across the disk. However, in this model approach, it is fit as a
constant and the value of the depletion factor doesn't affect the
derived value of  T$_{\rm CO}$ or the location of the CO snow line
R$_{\rm CO}$ in our model fit.
The bottom panel of Figure~\ref{fig:c18omodels}
shows the C$^{18}$O column density profile of the best-fit model
compared with the original \citet{Qi11} model.
In this model, the optical depth of $^{13}$CO $J=2-1$ is about unity
at around 155 AU (with $^{13}$CO column density around
7$\times$10$^{16}$ cm$^{-1}$ assuming 
$^{13}$CO/C$^{18}$O is around 8) , where marks the transition from 
optically thick to thin for $^{13}$CO $J=2-1$ emission.  
The match of the best-fit model with 
the C$^{18}$O data is much improved compared to the original \citet{Qi11} 
model (Fig. \ref{fig:c18omodels}).  The fit is not perfect, however; residuals in the inner disk suggest 
that the simple parametric form used for the CO abundance profile, and/or the 
adopted disk structure model, are imperfect at the smallest radial
scales.  

That said, the CO snow line location derived from the C$^{18}$O emission is in 
excellent agreement with the N$_2$H$^+$ inner ring edge.  This again confirms 
that this chemical signature in the N$_2$H$^+$ emission morphology provides a 
robust tracer of this major volatile snow line.  Moreover, the 
inferred discrepancy between the CO snow line locations derived from the
$^{13}$CO and C$^{18}$O emission highlights the potential dangers of using only 
CO isotopologues to constrain the CO snow line: rapid changes in line opacities 
can masquerade as snow lines.  This latter effect is especially problematic in 
cold disks where the CO snow line is close to the central star, as well as in 
very massive disks (as is the case here).

\subsection{The DCO$^+$ abundance profile}

\begin{figure*}[htp]
\epsscale{1}
\plotone{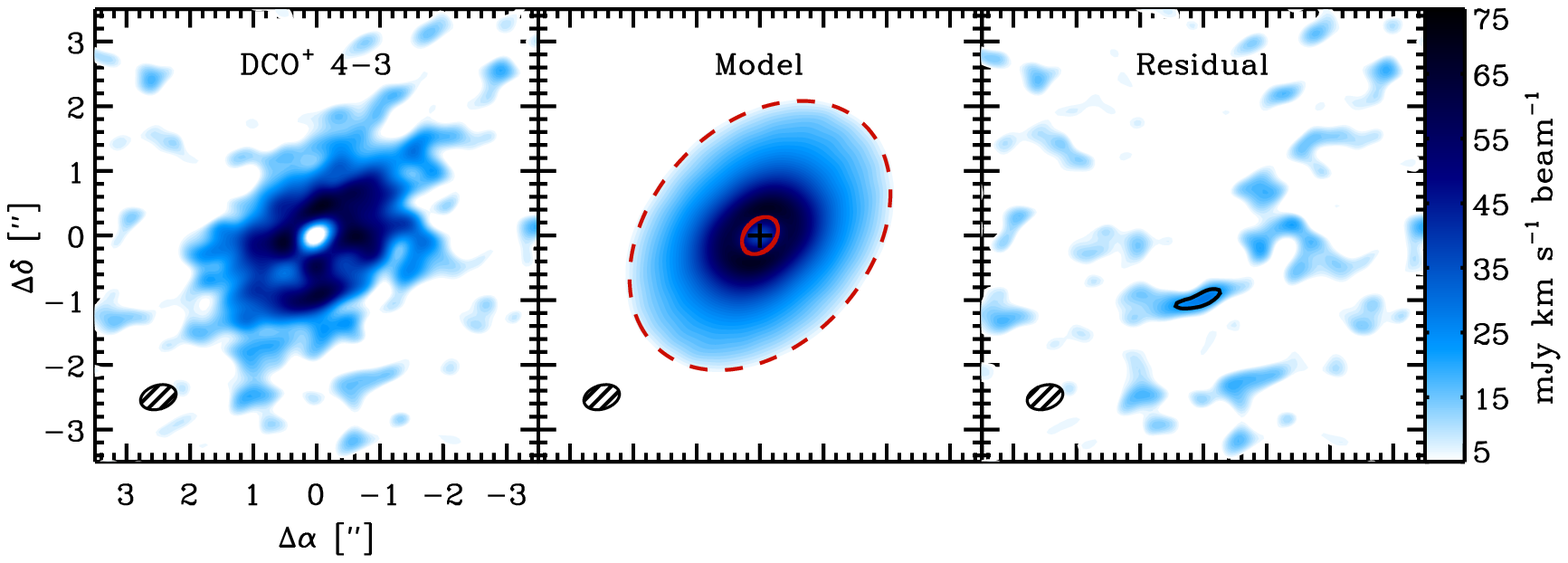}
\caption{ 
Integrated intensity image of DCO$^+$
  $4-3$, simulated image of the best-fit model, and the residual. The
  red solid and dashed ellipses mark the best-fit inner radius of 40
  AU and outer radius of 290 AU of the  DCO$^+$ ring. The residual
  shows also the contours in steps of 3$\sigma$. 
The integrated line
  emission scale is shown to the right of the upper right
  panel. Synthesized beams are drawn in the bottom left corner of each
  panel.  \label{fig:dcop43mom}} 
\end{figure*}

We also modeled the DCO$^+$~$J=4-3$ emission pattern following the same
strategy as for N$_2$H$^+$. We assumed a different set of vertical
boundaries, taking into account model predictions that the DCO$^+$
abundance should peak at elevated temperatures (above the CO
freeze-out temperature) compared to
N$_2$H$^+$.  However, as for N$_2$H$^+$, these have little effect on the inner 
and outer radial boundaries.  Figure~\ref{fig:dcop43mom} shows the best-fit 
model of DCO$^+$ and how it compares with the data. Similar to N$_2$H$^+$, the
best-fit model reproduces the data very well. Table~\ref{tab:models}
summarizes the best-fit parameters. As might be expected given 
Figure~\ref{fig:hdmaps}, both the inner and outer radial abundance boundaries 
are smaller for DCO$^+$ than for N$_2$H$^+$, 40 and 290~AU, respectively 
(compared to 90 and 500~AU for N$_2$H$^+$). Contrary to the expectations from 
models of simple chemical pathways, the outer DCO$^+$ emission boundary does 
not coincide with the derived CO snow line location of 90~AU (nor with
the previous $^{13}$CO-based snow line estimate of 155~AU).  

In addition, we explored the sensitivity of the derived DCO$^+$ radial 
abundance structure on excitation, employing ALMA Science Verification data for 
a higher energy transition, $J=5-4$ \citep[see][]{Mathews13}.  With the 
uncertainties, the derived best-fit model is consistent with what we find here 
for the $J$=4$-$3 transition (the inner boundary is $30\pm10$~AU and the outer 
boundary is $280\pm30$~AU).  Put simply, there is no obvious relationship 
between the observed radial distribution of DCO$^+$ emission and the location 
of the CO snow line.

\section{Discussion\label{sec:disc}}

\subsection{Chemical tracers of CO snow line}

N$_2$H$^+$, CO isotopologues, and DCO$^+$ have all been used to
chemically image the location of the CO snow line in disks
\citep{Qi11, Qi13c, Mathews13}. Our analysis of the distributions of
N$_2$H$^+$, $^{13}$CO, C$^{18}$O, and DCO$^+$ in the disk around
HD~163296 shows that the estimates of the CO snow line location from N$_2$H$^+$ 
and the optically thin isotopologue C$^{18}$O agree: both suggest that CO 
freeze-out starts at $\sim$90~AU (in the midplane).  By contrast, the utility 
of $^{13}$CO in this effort is limited by opacity effects. The DCO$^+$ emission 
does appear as a ring, and its overall emission structure must be related to CO 
freeze-out since one of its reactants is gas-phase CO.  However, neither the 
inner nor the outer DCO$^+$ emission edges coincide with the CO snow line 
location, suggesting a more complicated relationship between CO freeze-out and 
DCO$^+$ emission than was proposed by \citet{Mathews13}, perhaps
relating to the ``higher temperature'' chemical channel discussed by
\citet{Favre15}.   

These data indicate that both N$_2$H$^+$ and optically thin CO isotopologues 
can be used to trace the CO snow line at a disk midplane.  Of the two tracers, 
N$_2$H$^+$ provides, perhaps surprisingly, a more robust measure of the snow 
line location since its defining structure, the location of the inner edge
of the N$_2$H$^+$ ring, is relatively independent of model
assumptions and line opacities.

\subsection{CO freeze-out temperature and CO snow line}

The CO snow line location depends on both the disk
temperature profile and the CO freeze-out temperature. The latter is
set by a combination of ice physics and chemistry. The binding energy
of CO in pure CO ice is small, resulting in a CO freeze-out/desorption
temperature of $\sim$20~K. CO binds considerably stronger to H$_2$O
ice, which can raise the CO desorption temperature by up to 10~K
\citep{Collings04}. Whether CO will mainly bind to other CO molecules
or to H$_2$O molecules in the icy grain mantle depends on the chemical
history of the ice, i.e. whether H$_2$O and CO co-formed or
formed/condensed sequentially \citep[e.g.][]{Garrod11}.    

For the HD~163296 disk, we find a CO snow line location that corresponds to a
CO freeze-out temperature of 25~K. This is 8~K higher compared to  the
$T_{\rm CO}=$17 K found in the disk around TW~Hya \citep{Qi13c}. This
difference may be explained by the different chemical histories or
evolutionary stages of the two disks. TW~Hya is considerably older
than HD~163296, and its disk may therefore have experienced more chemical
processing that result in more segregated ices, where CO binds mainly
to other CO molecules. Higher CO freeze-out temperatures of about
  30 K found in embedded protostars \citep{Jorgensen15} also support
  the possible scenario of different ice compositions at different evolutionary
  stages. 
The difference in $T_{\rm CO}$ might also be due to 
different levels of UV ice processing. HD~163296 is a Herbig Ae
star, which produces a considerably more luminous UV radiation field compared 
to TW~Hya. This may have resulted in the substantial
production of complex organic molecules in the ices
\citep{Oberg11d}; much like for H$_2$O, this is expected to increase the
CO ice binding energies. A third potential cause is an overall
difference in the CO to H$_2$O abundance in the two disks (i.e., a lower
ratio in the HD~163296 disk). 

Understanding the origins of the observed difference in CO binding
energy is important to interpret Solar System observations, and to
predict the locations of CO snow lines in other disks. In the case of
a solar type host, a CO freeze-out temperature difference of 8~K is
equivalent to a difference of $\sim$10-20~AU in the CO snow line
location (i.e., the difference between Uranus and most comets forming
with or without CO). Resolving this question, and constraining how
common a high $T_{\rm CO}$ is in disks around solar-type hosts, is therefore 
key to determining the origin of high O and C abundances in ice giants in
the Solar System and to more generally assessing the likelihood of
planet formation around the CO snow line. This strongly
motivates the need for observational constraints on the distribution
of CO snow lines in larger samples of disks.  

\section{Conclusions}

We have presented and analyzed new  ALMA Cycle 1 observations of
N$_2$H$^+$ and DCO$^+$ 
emission, along with archival CO isotopologue data toward the disk around
HD~163296 with the aim to constrain the CO snow line location. Our key
results are:  

\begin{enumerate} 

\item Both the N$_2$H$^+$ $J=3-2$ and DCO$^+$ $J=4-3$ emission exhibit ring-shaped morphologies.  The best-fit models for the column density distribution of 
N$_2$H$^+$ has an inner cut-off at 90$^{+8}_{-6}$ AU, which we associate with 
the CO snow line location. The analogous best-fit model for DCO$^+$ has an 
inner abundance boundary at 40$^{+6}_{-3}$ AU and an outer boundary at 290$^{+6}_{-8}$ AU. 

\item Analysis of the C$^{18}$O $J=2-1$ data indicates a CO snow line
  location that is consistent with the N$_2$H$^+$ results, but not with
  the 155 AU snow line previously derived from $^{13}$CO data \citep{Qi11}. The
  agreement on snow line location when modeling N$_2$H$^+$ and
  optically thin CO isotopologue emission shows that N$_2$H$^+$ is a
  robust CO snow line tracer, while the relationship between DCO$^+$
  and CO freeze-out in the disk midplane is more complicated. 

\item The CO freeze-out temperature in the HD~163296 disk is $\sim$25
  K, which is considerably higher than the 17~K found for the disk around
  TW~Hya. This difference can be explained by different ice compositions
  and/or morphologies in the two disks. The difference also highlights
  the need for a survey of CO snow line radii, since their locations
  will be difficult to predict from theory without more observations. 

\end{enumerate}

{\it Facilities:} \facility{ALMA}

\acknowledgments

\noindent  
This paper makes use of the following ALMA data:
ADS/JAO.ALMA\#2012.1.00681.S and ADS/JAO.ALMA\#2011.0.00010.SV. ALMA
is a partnership of ESO 
(representing its member states), NSF (USA) and NINS (Japan), together
with NRC (Canada) and NSC and ASIAA (Taiwan), in cooperation with the
Republic of Chile. The Joint ALMA Observatory is operated by ESO,
AUI/NRAO and NAOJ. We acknowledge NASA Origins of Solar Systems grant
No. NNX11AK63. This paper is dedicated to the memory of Paola
D'Alessio, who was one of the co-Is of project 2012.1.00681.S
and passed away in November of 2013. 

\bibliographystyle{aa}

\begin{thebibliography}{32}
\expandafter\ifx\csname natexlab\endcsname\relax\def\natexlab#1{#1}\fi

\bibitem[{{Aikawa} \& {Nomura}(2006)}]{Aikawa06}
{Aikawa}, Y. \& {Nomura}, H. 2006, \apj, 642, 1152

\bibitem[{{Ali-Dib} {et~al.}(2014){Ali-Dib}, {Mousis}, {Petit}, \&
  {Lunine}}]{Ali-Dib14}
{Ali-Dib}, M., {Mousis}, O., {Petit}, J.-M., \& {Lunine}, J.~I. 2014, \apj,
  793, 9

\bibitem[{{Bailli{\'e}} {et~al.}(2015){Bailli{\'e}}, {Charnoz}, \&
  {Pantin}}]{Baillie15}
{Bailli{\'e}}, K., {Charnoz}, S., \& {Pantin}, E. 2015, \aap, 577, A65

\bibitem[{{Bergin} {et~al.}(2002){Bergin}, {Alves}, {Huard}, \&
  {Lada}}]{Bergin02}
{Bergin}, E.~A., {Alves}, J., {Huard}, T., \& {Lada}, C.~J. 2002, \apjl, 570,
  L101

\bibitem[{{Bergin} {et~al.}(2001){Bergin}, {Ciardi}, {Lada}, {Alves}, \&
  {Lada}}]{Bergin01}
{Bergin}, E.~A., {Ciardi}, D.~R., {Lada}, C.~J., {Alves}, J., \& {Lada}, E.~A.
  2001, \apj, 557, 209

\bibitem[{{Caselli} {et~al.}(1999){Caselli}, {Walmsley}, {Tafalla}, {Dore}, \&
  {Myers}}]{Caselli99}
{Caselli}, P., {Walmsley}, C.~M., {Tafalla}, M., {Dore}, L., \& {Myers}, P.~C.
  1999, \apjl, 523, L165

\bibitem[{{Chiang} \& {Youdin}(2010)}]{Chiang10}
{Chiang}, E. \& {Youdin}, A.~N. 2010, Annual Review of Earth and Planetary
  Sciences, 38, 493

\bibitem[{{Ciesla} \& {Cuzzi}(2006)}]{Ciesla06}
{Ciesla}, F.~J. \& {Cuzzi}, J.~N. 2006, \icarus, 181, 178

\bibitem[{{Collings} {et~al.}(2004){Collings}, {Anderson}, {Chen}, {Dever},
  {Viti}, {Williams}, \& {McCoustra}}]{Collings04}
{Collings}, M.~P., {Anderson}, M.~A., {Chen}, R., {et~al.} 2004, \mnras, 354,
  1133

\bibitem[{{Favre} {et~al.}(2015){Favre}, {Bergin}, {Cleeves}, {Hersant}, {Qi},
  \& {Aikawa}}]{Favre15}
{Favre}, C., {Bergin}, E.~A., {Cleeves}, L.~I., {et~al.} 2015, \apjl, 802, L23

\bibitem[{{Flower}(1999)}]{Flower99}
{Flower}, D.~R. 1999, \mnras, 305, 651

\bibitem[{{Garrod} \& {Pauly}(2011)}]{Garrod11}
{Garrod}, R.~T. \& {Pauly}, T. 2011, \apj, 735, 15

\bibitem[{{Gundlach} {et~al.}(2011){Gundlach}, {Kilias}, {Beitz}, \&
  {Blum}}]{Gundlach11}
{Gundlach}, B., {Kilias}, S., {Beitz}, E., \& {Blum}, J. 2011, \icarus, 214,
  717

\bibitem[{{Hogerheijde} \& {van der Tak}(2000)}]{Hogerheijde00}
{Hogerheijde}, M.~R. \& {van der Tak}, F.~F.~S. 2000, \aap, 362, 697

\bibitem[{{Johansen} {et~al.}(2007){Johansen}, {Oishi}, {Low}, {Klahr},
  {Henning}, \& {Youdin}}]{Johansen07}
{Johansen}, A., {Oishi}, J.~S., {Low}, M.-M.~M., {et~al.} 2007, \nat, 448, 1022

\bibitem[{{J{\o}rgensen}(2004)}]{Jorgensen04}
{J{\o}rgensen}, J.~K. 2004, \aap, 424, 589

\bibitem[J{\o}rgensen et 
al.(2015)]{Jorgensen15} J{\o}rgensen, J.~K., Visser, R., Williams, J.~P., \& Bergin, E.~A.\ 2015, \aap, 579, A23 

\bibitem[Loomis et al.(2015)]{Loomis15} Loomis, R.~A., Cleeves, 
L.~I., {\"O}berg, K.~I., Guzman, V.~V., 
\& Andrews, S.~M.\ 2015, \apjl, 809, L25 


\bibitem[{{Mathews} {et~al.}(2013){Mathews}, {Klaassen}, {Juh{\'a}sz},
  {Harsono}, {Chapillon}, {van Dishoeck}, {Espada}, {de Gregorio-Monsalvo},
  {Hales}, {Hogerheijde}, {Mottram}, {Rawlings}, {Takahashi}, \&
  {Testi}}]{Mathews13}
{Mathews}, G.~S., {Klaassen}, P.~D., {Juh{\'a}sz}, A., {et~al.} 2013, \aap,
  557, A132

\bibitem[{{Mumma} \& {Charnley}(2011)}]{Mumma11}
{Mumma}, M.~J. \& {Charnley}, S.~B. 2011, \araa, 49, 471

\bibitem[{{{\"O}berg} {et~al.}(2011{\natexlab{a}}){{\"O}berg}, {Boogert},
  {Pontoppidan}, {van den Broek}, {van Dishoeck}, {Bottinelli}, {Blake}, \&
  {Evans}}]{Oberg11c}
{{\"O}berg}, K.~I., {Boogert}, A.~C.~A., {Pontoppidan}, K.~M., {et~al.}
  2011{\natexlab{a}}, \apj, 740, 109

\bibitem[{{{\"O}berg} {et~al.}(2011{\natexlab{b}}){{\"O}berg}, {Qi}, {Wilner},
  \& {Andrews}}]{Oberg11d}
{{\"O}berg}, K.~I., {Qi}, C., {Wilner}, D.~J., \& {Andrews}, S.~M.
  2011{\natexlab{b}}, \apj, 743, 152

\bibitem[{{{\"O}berg} {et~al.}(2011{\natexlab{c}}){{\"O}berg}, {van der Marel},
  {Kristensen}, \& {van Dishoeck}}]{Oberg11b}
{{\"O}berg}, K.~I., {van der Marel}, N., {Kristensen}, L.~E., \& {van
  Dishoeck}, E.~F. 2011{\natexlab{c}}, \apj, 740, 14

\bibitem[{{{\"O}berg} {et~al.}(2008){{\"O}berg}, {van Dishoeck}, \&
  {Linnartz}}]{Oberg08b}
{{\"O}berg}, K.~I., {van Dishoeck}, E.~F., \& {Linnartz}, H. 2008, in IAU
  Symposium, Vol. 251, IAU Symposium, ed. S.~{Kwok} \& S.~{Sanford}, 449--450

\bibitem[{{Oka} {et~al.}(2012){Oka}, {Inoue}, {Nakamoto}, \& {Honda}}]{Oka12}
{Oka}, A., {Inoue}, A.~K., {Nakamoto}, T., \& {Honda}, M. 2012, \apj, 747, 138

\bibitem[{{Qi} {et~al.}(2011){Qi}, {D'Alessio}, {{\"O}berg}, {Wilner},
  {Hughes}, {Andrews}, \& {Ayala}}]{Qi11}
{Qi}, C., {D'Alessio}, P., {{\"O}berg}, K.~I., {et~al.} 2011, \apj, 740, 84

\bibitem[{{Qi} {et~al.}(2013{\natexlab{a}}){Qi}, {{\"O}berg}, \&
  {Wilner}}]{Qi13a}
{Qi}, C., {{\"O}berg}, K.~I., \& {Wilner}, D.~J. 2013{\natexlab{a}}, \apj, 765,
  34

\bibitem[{{Qi} {et~al.}(2013{\natexlab{b}}){Qi}, {{\"O}berg}, {Wilner},
  {D'Alessio}, {Bergin}, {Andrews}, {Blake}, {Hogerheijde}, \& {van
  Dishoeck}}]{Qi13c}
{Qi}, C., {{\"O}berg}, K.~I., {Wilner}, D.~J., {et~al.} 2013{\natexlab{b}},
  Science, 341, 630

\bibitem[{{Ros} \& {Johansen}(2013)}]{Ros13}
{Ros}, K. \& {Johansen}, A. 2013, \aap, 552, A137

\bibitem[{{Rosenfeld} {et~al.}(2013){Rosenfeld}, {Andrews}, {Hughes}, {Wilner},
  \& {Qi}}]{Rosenfeld13}
{Rosenfeld}, K.~A., {Andrews}, S.~M., {Hughes}, A.~M., {Wilner}, D.~J., \&
  {Qi}, C. 2013, \apj, 774, 16

\bibitem[{{Sch{\"o}ier} {et~al.}(2005){Sch{\"o}ier}, {van der Tak}, {van
  Dishoeck}, \& {Black}}]{Schoier05}
{Sch{\"o}ier}, F.~L., {van der Tak}, F.~F.~S., {van Dishoeck}, E.~F., \&
  {Black}, J.~H. 2005, \aap, 432, 369

\bibitem[Walsh et al.(2010)]{Walsh10} Walsh, C., Millar, T.~J., 
\& Nomura, H.\ 2010, \apj, 722, 1607 

\bibitem[Willacy(2007)]{Willacy07} Willacy, K.\ 2007, \apj, 660, 
441 

\end{thebibliography}

\appendix
\section{Channel images}

Figure~\ref{fig:n2hpchannelmaps} and~\ref{fig:dcop43channelmaps} show
channel maps for N$_2$H$^+$ $J=3-2$ and DCO$^+$ $J=4-3$ and their
best-fit models along with the residuals. The best-fit models are
described in Table~\ref{tab:models}.

\begin{figure*}[htbp]
\centering
\includegraphics[width=5.3in]{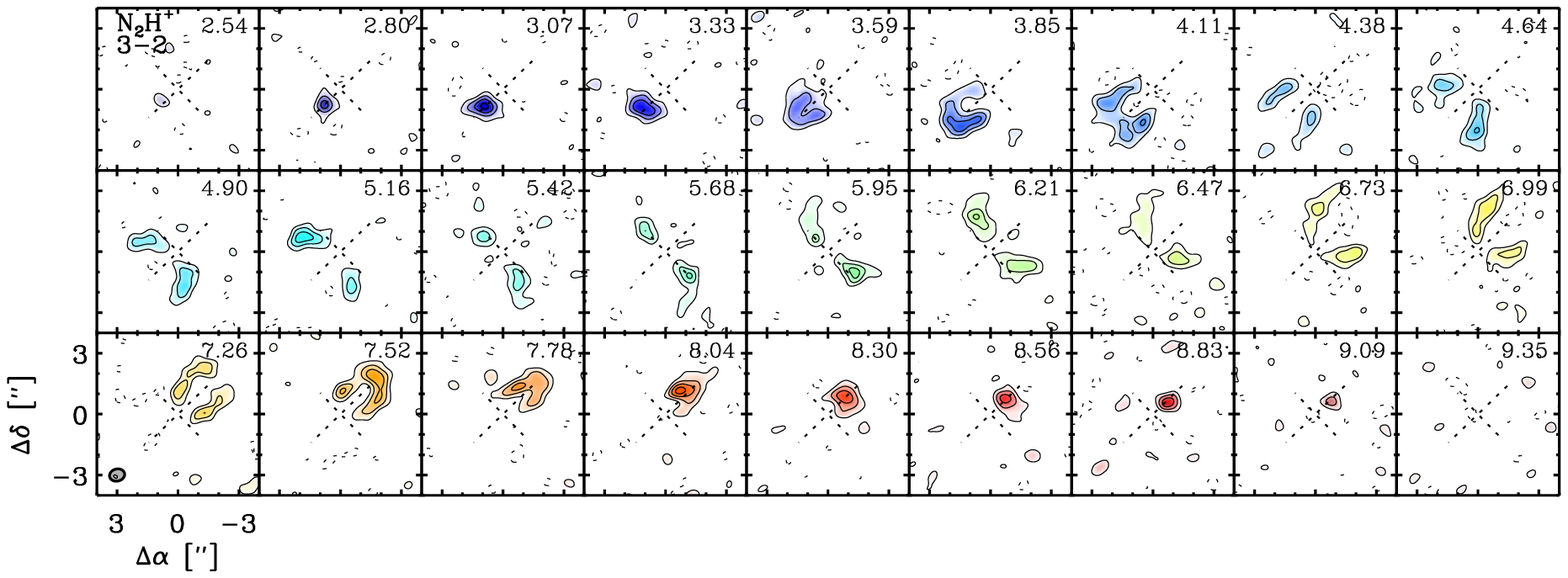}
\vskip 5mm
\includegraphics[width=5.3in]{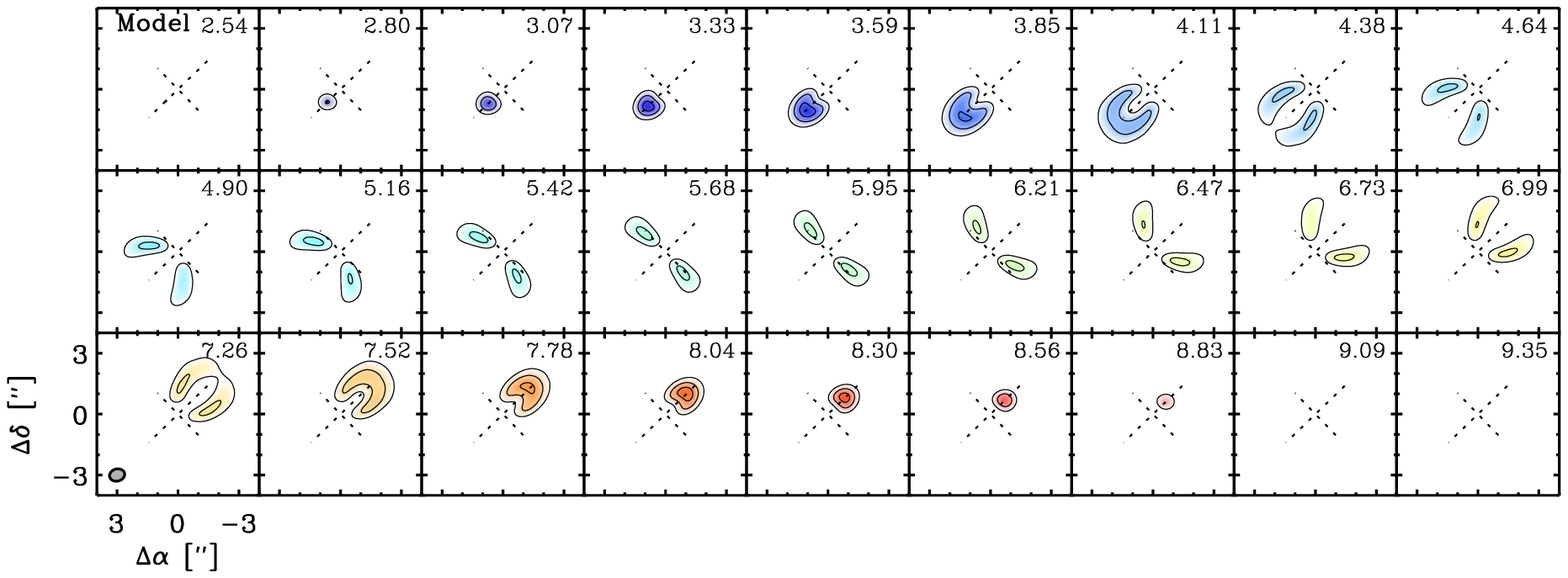}
\vskip 5mm
\includegraphics[width=5.3in]{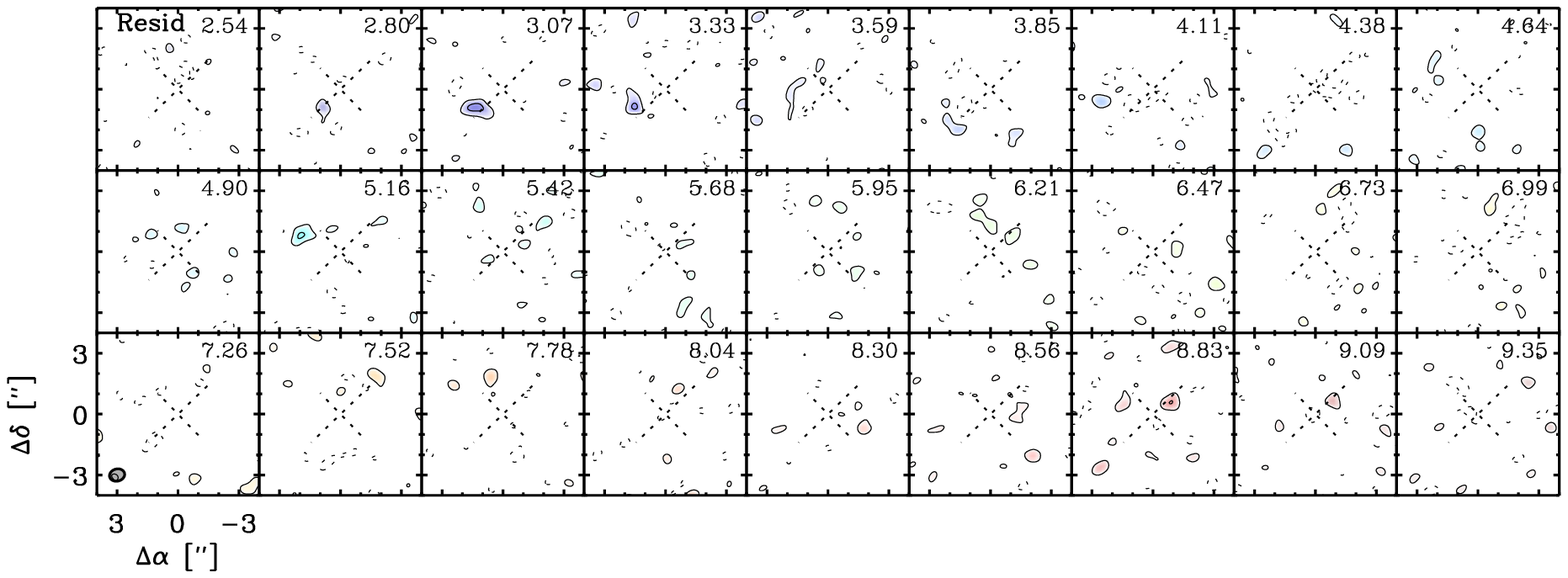}
\caption{ Channel maps of the N$_2$H$^+$ $J=3-2$ line emission toward
  the disk of HD~163296. The LSR velocity is indicated in the upper right
  of each channel, while the synthesized beam size and orientation
  ($0.''75 \times 0.''60$ at a position angle of $-$75.4$^{\circ}$)
  is indicated in the lower left  
  panel. The contours are 0.0035 (1$\sigma$) $\times
  [2,4,6,8,10,12]$ Jy beam$^{-1}$ .}
 \label{fig:n2hpchannelmaps}
\end{figure*}

\begin{figure*}[htbp]
\centering
\includegraphics[width=5.3in]{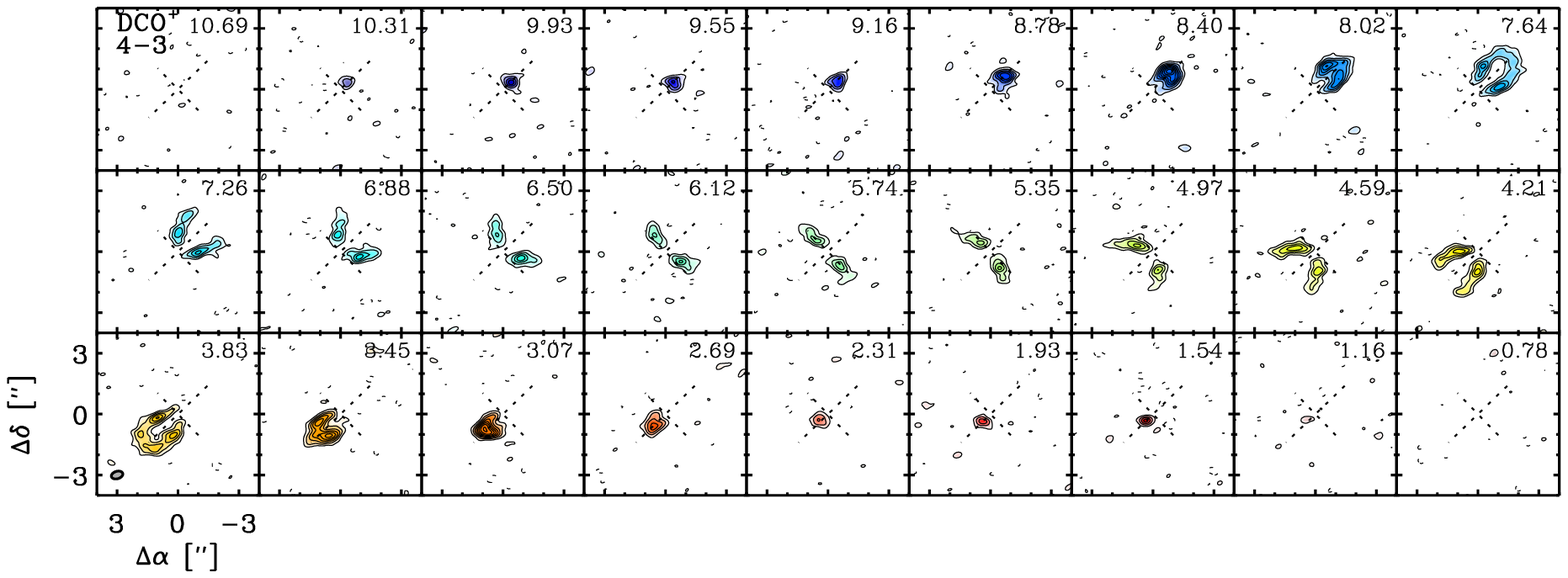}
\vskip 5mm
\includegraphics[width=5.3in]{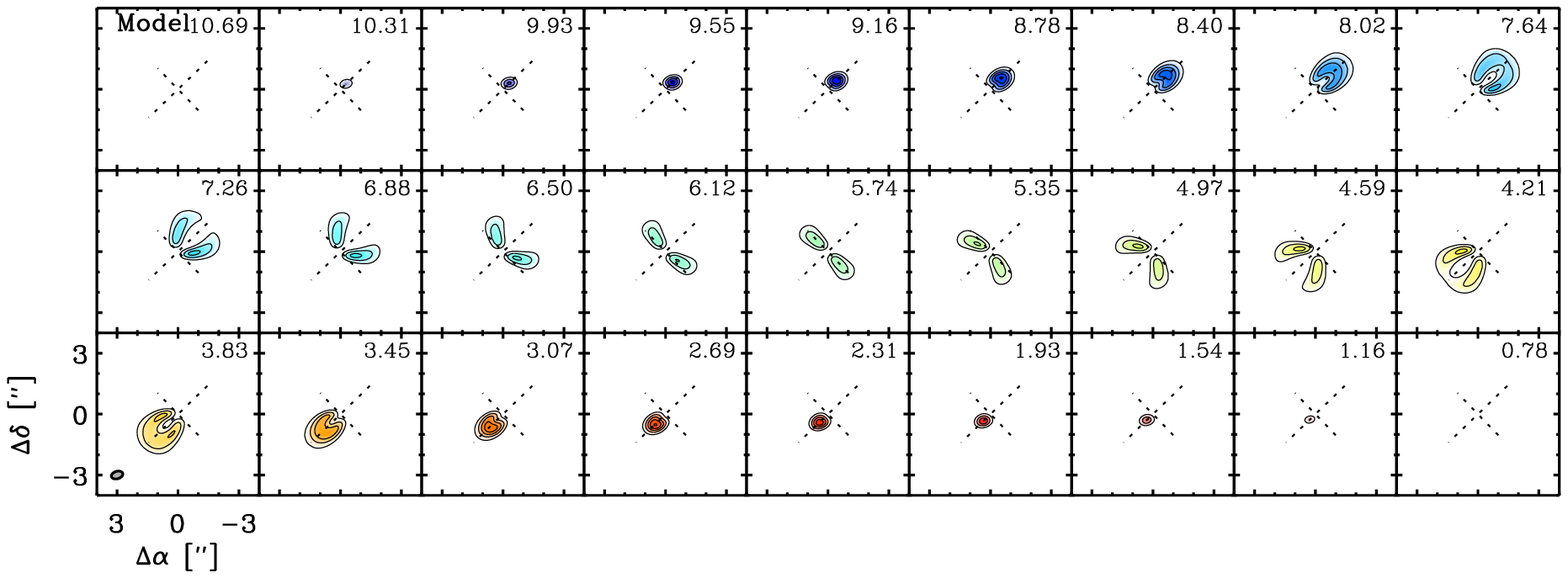}
\vskip 5mm
\includegraphics[width=5.3in]{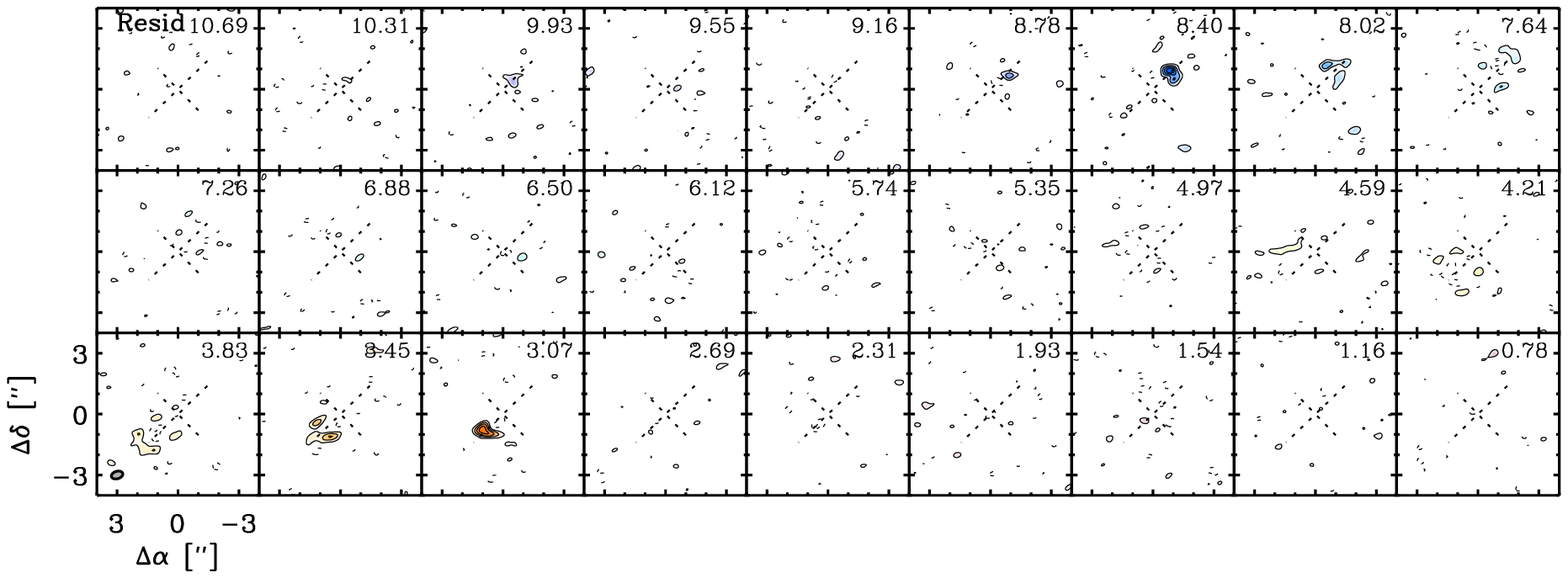}
\caption{ Same as Fig~\ref{fig:n2hpchannelmaps} but for DCO$^+$
  $J=4-3$ line emission and the synthesized beam size and orientation
  ($0.''58 \times 0.''37$ at a position angle of $-$73.2$^{\circ}$)
  is indicated in the lower left  
  panel. The contours are 0.04 (1$\sigma$) $\times
  [2,4,6,8,10,12,14,16]$ Jy beam$^{-1}$ .}
 \label{fig:dcop43channelmaps}
\end{figure*}

\end{document}